\documentstyle [fleqn,12pt] {article}
\def\mscript#1{\mbox{\scriptsize$#1$}}

\begin{document}
\pagestyle{empty}
\begin{flushleft}
\end{flushleft}
\vspace*{2cm}
\begin{center}
{\bf Thermofield Dynamics of the Heterotic 
String \\
 --- Physical Aspects of the Thermal Duality --- } \\
\vspace{1cm} 
H. Fujisaki
\vspace{5mm}

{\it Department of Physics, Rikkyo University, Tokyo 171}\\
\vspace*{3cm}
{\bf ABSTRACT}
\end{center}

\indent The thermofield dynamics of the $D = 10$ 
heterotic thermal string theory is described in proper reference to 
the thermal duality symmetry as well as the 
thermal stability of modular invariance in association with the 
global phase structure of the heterotic thermal string 
ensemble.    
\newpage
\pagestyle{plain}
\setcounter{page}{1}
\indent Building up thermal string theories based upon the thermofield 
dynamics (TFD) \cite{umezawa} has gradually been endeavoured in leaps 
and bounds \cite{leblanc1} -- \cite{nakagawa}.   
In the present 
communication, physical aspects of the thermal duality symmetry 
\cite{obrien} -- \cite{osorio} are 
commentarially  
exemplified $\grave{a}\: la$ recent publication of ourselves \cite{fujisaki4} 
through the infrared 
behaviour of the one-loop cosmological constant in proper respect of 
the thermal stability of modular invariance for the $D = 10$ 
heterotic thermal string theory based upon the TFD algorithm.  The 
global phase structure of the $D = 10$ heterotic thermal string ensemble is
then 
recapitulatively touched upon.  

Let us start with the one-loop cosmological constant 
$\Lambda(\beta)$ as follows:
\vspace{5mm}
\begin{equation}
\Lambda (\beta) = \frac{\alpha^\prime}{2} \lim_{\mu^2 \rightarrow 0} {\rm 
Tr} 
\left[ \int_{\infty}^{\mu^2} dm^2 \left( \Delta^\beta_B (p, P; m^2) + 
\Delta^{\beta}_{F} (p, P; m^2) \right) \right]
\end{equation}

\vspace{5mm}
\noindent at any finite temperature $\beta^{-1} = kT$ in the $D = 10$ heterotic
 thermal string theory based upon the TFD algorithm, 
where $\alpha^\prime$ means the slope parameter, $p^\mu$ 
reads loop momentum, $P^I$ lie on the even self-dual root lattice $L = 
\Gamma_8 \times \Gamma_8$ for the exceptional group $G = E_8 \times 
E_8$ \cite{green} and the thermal propagator 
$\Delta^\beta_{B[F]} (p, P; m^2)$ of the free closed bosonic [fermionic] 
string is written $\grave{a}\: la$ Leblanc \cite{leblanc1} in the form
 
\begin{eqnarray}
\lefteqn{\Delta^\beta_{B[F]}(p, P; m^2) = \int_{-\pi}^{\pi} 
\frac{d\phi}{4\pi} \; {\rm e}^{i\phi \left( N - \alpha - \bar{N} + 
\bar{\alpha} -1/2 \cdot \sum_{I=1}^{16} (P^I)^2 \right) }}
\nonumber \\
& & \times \Biggl( \left[ \raisebox{-1ex}{$\stackrel {\textstyle 
+}{\mscript{[}-\mscript{]}}$} \int_{0}^{1} dx + \frac{1}{2}
\sum_{n=0}^{\infty} \frac{\delta [\alpha^\prime 
/2 \cdot p^2 + \alpha^\prime /2 \cdot m^2 + 2(n - \alpha)]}{{\rm e}^{\beta 
|p_0|} \raisebox{-1ex}{$\stackrel{\textstyle -}{\mscript{[}+\mscript{]}}$}
\; 1} 
\oint_{c} dx \right] \nonumber \\
& & \times x^{\alpha^\prime /2 \cdot p^2 + N - \alpha + \bar{N} - 
\bar{\alpha} + 1/2 \cdot \sum_{I=1}^{16} (P^I)^2 + \alpha^\prime /2 \cdot 
m^2 - 1} \Biggr) \quad ,
\end{eqnarray}

\vspace{5mm}
\noindent where $N$ [$\bar{N}$] denotes the number operator of the right- 
[left-] 
mover, the intercept parameter $\alpha$ [$\bar{\alpha}$] is fixed at
$\alpha = 0$  
$[\bar{\alpha} = 1]$ and the contour $c$ is taken as the unit circle 
around the origin.  We are then eventually led to 
the modular parameter integral representation of $\Lambda (\beta)$ at $D = 10$ 
as follows \cite{fujisaki4}:
\vspace{5mm}
\begin{eqnarray}
\Lambda (\beta) & = & -8(2\pi \alpha^\prime)^{-D/2} \int_{E} 
\frac{d^2\tau}{2\pi \tau_2^2} \; (2\pi \tau_2)^{-(D-2)/2}\; {\rm 
e}^{2\pi i \bar{\tau}} \left[ 1 + 480 \sum_{m=1}^{\infty} 
\sigma_7(m)\bar{z}^m \right] \nonumber \\
& & \times \prod_{n=1}^{\infty} (1 - \bar{z}^n)^{-D-14} \left( 
\frac{1 + z^n}{1 - z^n} \right)^{D-2} \sum_{\ell \in Z; {\rm odd}} \exp \left[ 
- \frac{\beta^2}{4\pi \alpha^\prime \tau_2}\; \ell^2 \right] 
\rule{0cm}{1cm} \quad , 
\end{eqnarray}

\vspace{5mm}
\noindent where $\stackrel{[\normalsize{-}]}{\tau} = \tau_1 
\raisebox{-1ex}{$\stackrel{\normalsize{+}}{\mscript{[}\!-\!\mscript{]}}$} 
i\tau_2$, 
$z = x {\rm e}^{i\phi} = {\rm 
e}^{2\pi i \tau}$, $\bar{z} = x{\rm e}^{-i \phi} = {\rm e}^{-2\pi i 
\bar{\tau}}$, $E$ means the half-strip integration 
region in the complex $\tau$ plane, {\it i.e.} $-1/2 \leq \tau_1 \leq 1/2$; 
$\tau_2 > 0$.  Accordingly, the $D = 10$ 
thermal amplitude $\beta \Lambda (\beta)$ is identical  
with the ``$E$-type'' representation of the thermo-partition function 
$\Omega_h (\beta)$ of the heterotic string in ref.~\cite{obrien}.  
The ``$E$-type'' thermal amplitude $\Lambda (\beta)$ 
is not modular invariant and annoyed with ultraviolet 
divergences for $\beta \leq 
\beta_H = (2 + \sqrt{2}) \pi \sqrt{\alpha^\prime}$, where $\beta_H$ 
reads the inverse Hagedorn temperature of the heterotic thermal 
string.  Let us pay attention to the fact that the thermal amplitude 
$\Lambda (\beta)$ is infrared convergent for any value of $\beta$.     

Our prime concern is reduced to regularizing the thermal amplitude 
$\Lambda (\beta)$ {\it $\grave{a}$ la} ref.~\cite{obrien} as well as 
ref.~\cite{osorio} through transforming the physical information in 
the ultraviolet region of the half-strip $E$ into the ``new-fashioned'' 
modular invariant amplitude.  Let us postulate $\grave{a}\: la$ 
ref.~\cite{fujisaki4} 
the one-loop dual symmetric 
thermal cosmological 
constant $\bar{\Lambda} (\beta; D)$ at any space-time dimension $D$  as an 
integral over the 
fundamental domain $F$, {\it i.e.} $-1/2 
\leq \tau_1 \leq 1/2 \: ; \tau_2 > 0 \: ; |\tau| > 1$, of the modular 
group 
$SL(2, Z)$ as follows:

\begin{eqnarray}
\bar{\Lambda} (\beta; D) & = & \frac{2}{\beta} (2\pi \alpha^\prime)^{-D/2} 
\sum_{(\sigma, \rho)} 
\int_{F} 
\frac{d^2\tau}{2\pi \tau_2^2}\; (2 \pi \tau_2)^{-(D-2)/2}\;  
\bar{z}^{-(D+14)/24} z^{-(D-2)/24}  \nonumber \\
& & \times \left[ 1 + 480 \sum_{m=1}^{\infty} 
\sigma_7 (m) \bar{z}^m \right] \prod_{n=1}^{\infty} (1 - \bar{z}^n)^{-D-14} 
(1 - z^n)^{-D+2} \nonumber \\[5mm] 
& & \times A_{\sigma \rho} (\tau; D) \left[C_\sigma^{(+)}(\bar{\tau}, \tau;
\beta) 
+ \rho C_\sigma^{(-)}(\bar{\tau}, \tau; \beta) \right] \quad ,
\end{eqnarray}

where
\begin{eqnarray}
\left( \begin{array}{c}
A_{+-}(\tau; D) \rule[-2mm]{0mm}{8mm} \\ A_{-+}(\tau; D)
\rule[-2mm]{0mm}{8mm} \\ A_{--}(\tau; D) 
\end{array} \right) 
= 8 \left( \frac{\pi}{4} \right) ^{(D-2)/6} 
\left( \begin{array}{l}
-[\theta_2(0, \tau)/\theta_{1}^{\prime}(0, \tau)^{1/3}]^{(D-2)/2}
\rule[-2mm]{0mm}{8mm} \\
-[\theta_4(0, \tau)/\theta_{1}^{\prime}(0, \tau)^{1/3}]^{(D-2)/2}
\rule[-2mm]{0mm}{8mm} \\
\; [\theta_3(0, \tau)/\theta_{1}^{\prime}(0, \tau)^{1/3}]^{(D-2)/2}
\rule[-2mm]{0mm}{8mm}
\end{array} \right) \quad ,
\end{eqnarray}

\vspace{5mm}
\begin{equation}
C_\sigma^{(\gamma)}(\bar{\tau}, \tau; \beta) = (4\pi^2\alpha^\prime 
\tau_2)^{1/2} \sum_{(p, q)} \exp \left[ - \frac{\pi}{2} \left( 
\frac{\beta^2}{2\pi^2\alpha^\prime} p^2 + \frac{2\pi^2\alpha^\prime}
{\beta^2} q^2 \right) \tau_2 + i\pi pq \tau_1 \right] ,
\end{equation}

\vspace{5mm}
\noindent the signatures $\sigma, \rho$ and $\gamma$ read $\sigma, \rho = +, 
- ; \; -, + ; \; -, -$ and $\gamma = +, -$, respectively, and the summation 
over $p \; [q]$ is restricted by $(-1)^p = \sigma \; [(-1)^q = 
\gamma]$.   
It is almost needless to mention that the $D = 10$ thermal amplitude 
$\beta\bar{\Lambda}(\beta; D = 10)$ is literally reduced to the 
``$D$-type'' representation of the thermo-partition function 
$\Omega_h(\beta)$ in ref.~\cite{obrien} which in turn guarantees 
$\bar{\Lambda}(\beta; D = 10) = \Lambda (\beta)$  at least below the 
Hagedorn temperature $\beta_H^{-1}$.  Typical theoretical observations are
as follows 
\cite{fujisaki4}:  First of all, the thermal amplitude 
$\bar{\Lambda}(\beta; D)$ is manifestly modular invariant and 
free of ultraviolet divergences for any value of $\beta$ and $D$.    
If and only if $D =  10$, in addition, the thermal duality relation $\beta 
\bar{\Lambda}(\beta; D) = 
\tilde{\beta}\bar{\Lambda}(\tilde{\beta}; D)$ is manifestly satisfied 
for the thermal amplitude $\bar{\Lambda}(\beta; D)$, irrespective of 
the value of $\beta$, where $\tilde{\beta} = 2\pi^2 \alpha^\prime 
/\beta$.  By way of parenthesis, let us remember that the so-called 
thermal duality relation is not universal for closed string theories 
in general \cite{obrien} -- \cite{osorio} in sharp contrast to 
modular invariance.     

The infrared behaviour of the 
thermal cosmological constant $\bar{\Lambda}(\beta; D)$ 
is asymptotically described as \cite{fujisaki4}
\begin{eqnarray}
\bar{\Lambda}(\beta; D) & = & - 64 \sqrt{2}\: (8\pi^2 \alpha^\prime)^{-D/2} 
\sum_{(p, q)} \int_{- \frac{1}{2}}^{\frac{1}{2}} d\tau_1 \exp [i\pi 
pq\tau_1] \, \sqrt{\frac{\tilde{\beta}}{\beta}} 
\int_{\sqrt{1-\tau_1^2}}^{\infty} d\tau_2 \; \tau_2^{-(D+1)/2} \nonumber \\
& & \times \exp \left[ - \frac{\pi}{2} \; \tau_2 \left( 
\frac{\beta}{\tilde{\beta}} \; p^2 + \frac{\tilde{\beta}}{\beta} \; q^2 - 
\frac{5}{12} (D - 10) - 6 \right) \right] \quad ,
\end{eqnarray}

\vspace{5mm}
\noindent where $p, q = \pm 1; \pm 3; \pm 5; \cdots$. The $D = 10$ TFD
amplitude 
$\bar{\Lambda}(\beta; D = 10)$ is then infrared divergent for $(2 - \sqrt{2})
\pi \sqrt{\alpha^{\prime}} = \tilde{\beta}_H \leq \beta \leq \beta_H$ in 
association with the presence of the tachyonic mode, where 
$\tilde{\beta}_H$ reads the inverse dual Hagedorn temperature of the 
heterotic thermal string.   
We can therefore define $\grave{a}\: la$ ref.~\cite{fujisaki4} 
the dimensionally regularized, 
$D = 10$ one-loop dual symmetric thermal cosmological constant 
$\hat{\Lambda}(\beta)$ in the sense of analytic continuation from $D < 2/5$ 
to higher values of $D$, {\it i.e.} $D = 10$ by 
\begin{eqnarray}
\hat{\Lambda}(\beta) & = &  -\frac{2}{\beta} (8\pi 
\alpha^\prime)^{-(D-1)/2} \sum_{(p, q)} 
\int_{- \frac{1}{2}}^{\frac{1}{2}} d\tau_1 \exp[i\pi pq \tau_1] \nonumber 
\\
& & \times \left( \frac{\beta^2}{2\pi^2\alpha^\prime} \: p^2 + 
\frac{2\pi^2\alpha^\prime}
{\beta^2} \: q^2 - 6 - i\varepsilon \right) ^{(D-1)/2} \nonumber \\[5mm]
& & \times \Gamma \left[ - \frac{D - 1}{2}\: ,\; \frac{\pi}{2} \sqrt{1 - 
\tau_1^2} \left( \frac{\beta^2}{2\pi^2 \alpha^\prime} \: p^2 + 
\frac{2\pi^2 
\alpha^\prime}{\beta^2} \: q^2 - 6 - i\varepsilon \right) \right]\; ;\quad
D = 10 , 
\nonumber \\
&  & 
\end{eqnarray}

\vspace{5mm}
\noindent irrespective of the value of $\beta$, where $\Gamma$ is the 
incomplete gamma function of the second kind and the so-called $D + 
i\varepsilon$ procedure is to be adopted in natural consonance with the 
nonvanishing decay rate of the tachyonic thermal vacuum.  

The dimensionally 
regularized, thermal cosmological constant $\hat{\Lambda} (\beta)$
 manifestly satisfies the thermal duality relation $\beta 
\hat{\Lambda}(\beta) = \tilde{\beta} \hat{\Lambda}(\tilde{\beta})$ in 
full accordance with the thermal stability of modular invariance.  The 
thermal duality symmetry mentioned above immediately yields the 
asymptotic formula as follows \cite{fujisaki1}, \cite{atick}:
\begin{equation}
\lim_{\beta \rightarrow 0} \beta \hat{\Lambda} (\beta) = 
\lim_{\tilde{\beta}^{-1} \rightarrow 0} \tilde{\beta} 
\hat{\Lambda} (\tilde{\beta}) \; , 
\end{equation}

\noindent or equivalently
\begin{equation}
\hat{\Lambda} (\beta \sim 0) \sim \frac{2\pi^2 
\alpha^\prime}{\beta^2} \hat{\Lambda} (\beta^{-1} \rightarrow 0) = 
\frac{2\pi^2 \alpha^\prime}{\beta^2} \Lambda
\end{equation}

\noindent for the $D = 10$ heterotic thermal string theory, where $\Lambda$ 
literally reads the $D = 10$ zero-temperature, one-loop cosmological 
constant which is in turn guaranteed to vanish automatically as an 
inevitable consequence of the Jacobi identity $\theta_2^4 - \theta_3^4 + 
\theta_4^4 = 0$ for the theta functions.  Let us call to our 
remembrance that the vanishing machinery of the $D = 10$ 
zero-temperature amplitude $\Lambda$ is self-evident in the present 
context due to the absence of the term with $\ell = 0$ [$p = q = 0$] 
on the right hand side of eq.~(3) [eq.~(7) or equivalently eq.~(8)].  
The present observation is 
paraphrased $\grave{a}\: la$ ref.~\cite{osorio} as follows:  The 
thermal duality symmetry is inherent to the fact that the total number 
of degrees of freedom vanishes at extremely high temperature $\beta 
\sim 0$ in the sense of the modular invariant counting.  Accordingly, 
it seems possible to claim $\grave{a}\: la$ ref.~\cite{atick} that 
the $D = 10$ heterotic thermal string has no fundamental gauge 
invariant degrees of freedom at least at $\beta \sim 0$ and will be 
asymptotically described at high temperature by underlying 
topological theory. The present view may be in essential agreement 
with the provocative argument of Witten \cite{witten} on the possible 
unbroken high-energy phase of string theory in topological $\sigma$ 
models and will deserve more than passing consideration in an attempt 
to substantialize the crucial geometrical ideas purely topological in 
character, {\it e.g.} the possible background independence at 
asymptotically high energies in string theory.   
It is parenthetically mentioned that another 
newfangled hypothetical view might not be exhaustively excluded yet 
at least as a matter of taste in which $\bar{\Lambda}(\beta; D)$ and 
consequently 
$\hat{\Lambda}(\beta)$ would not be physically well-defined beyond 
the Hagedorn temperature $\beta_H^{-1}$.  

Let us describe $\grave{a}\: la$ ref.~\cite{fujisaki4} the singularity 
structure of the dimensionally 
regularized, dual symmetric thermal amplitude 
$\hat{\Lambda}(\beta)$.  The position of the singularity $\beta_{|p|, 
|q|}$ is determined by solving $\beta/\tilde{\beta} \cdot p^2 + 
\tilde{\beta}/\beta \cdot q^2 - 6 = 0$ for every allowed $(p, q)$ in 
eq.~(8).  Thus we obtain a set of solutions 
as follows: $\beta_{1, 1} = \beta_H = (\sqrt{2} + 
1) \pi \sqrt{2\alpha^\prime}$ and $\tilde{\beta}_{1, 1} = \tilde{\beta}_H 
= 
(\sqrt{2} - 1)\pi \sqrt{2\alpha^\prime}\:$ which form the leading 
branch points of the square root type at $\beta_H$ and $\tilde{\beta}_H$, 
respectively.  It is of 
practical significance to note that there exists no self-dual leading 
branch point at $\beta_0 = \tilde{\beta}_0 = \pi 
\sqrt{2\alpha^\prime}$. As a matter of fact, the regularized thermal 
amplitude $\hat{\Lambda}(\beta)$ develops the imaginary part across 
the leading branch cut $\tilde{\beta}_H \leq \beta \leq \beta_H$ in 
association with instability of the tachyonic thermal vacuum.   
We are now in the position to touch upon {\it 
$\grave{a}$ la} ref.~\cite{fujisaki4} the global 
phase structure of the $D = 10$ heterotic thermal string ensemble.  
There 
will then exist three phases in the sense of the thermal duality 
symmetry as follows \cite{fujisaki4}, \cite{obrien}, \cite{leblanc3}: (i) 
the $\beta$ channel canonical phase in the tachyon-free region $(2 +
\sqrt{2})\pi \sqrt
{\alpha^\prime} = \beta_H \leq \beta < \infty$, (ii) the dual 
$\tilde{\beta}$ channel canonical phase in the tachyon-free region 
$0 < \beta \leq 
\tilde{\beta}_H = (2 - \sqrt{2})\pi \sqrt{\alpha^\prime}$ and 
(iii) the self-dual microcanonical phase in the tachyonic region 
$\tilde{\beta}_H < 
\beta 
< \beta_H$.  
There 
will appear no effective splitting of the microcanonical domain because of 
the absence of the self-dual branch point at $\beta_0 = \tilde{\beta}_0 
= \pi \sqrt{2\alpha^\prime}$.  As a 
consequence,  it still remains to be clarified whether the so-called maximum 
temperature of 
the heterotic string excitation is asymptotically described as
 $\beta_0^{-1} = \tilde{\beta}_0^{-1}$ in proper respect of the
self-duality of the microcanonical 
 phase.  If the thermal duality relation were tentatively supposed to 
 be manifestly broken, by way of parenthesis, there would then appear 
 the essential singularity at the infinite temperature $\beta = 0$ as 
 the accumulation point of infinitely many finite-temperature branch 
 points beyond the Hagedorn temperature $\beta_H^{-1}$ 
 \cite{leblanc2}, \cite{fujisaki1}, \cite{fujisaki3}. 
  
We have succeeded in shedding some light upon the global phase 
structure of the thermal string ensemble through the infrared 
behaviour of the one-loop free energy amplitude for the dimensionally 
regularized, $D = 10$ heterotic thermal string theory based upon the 
TFD algorithm.  In particular, physical aspects of the thermal 
duality symmetry have been described in full harmony with the thermal
stability of modular 
invariance not only for the canonical region but also for 
the microcanonical region.  It is hoped that we can illuminate the fruitful 
thermodynamical investigation \cite{deo} of string excitations, {\it e.g.}
the manifest 
materialization of the ``true'' maximum temperature for the thermal 
string ensemble in general within the new-fashioned duality framework of the
 $D$-brane paradigm \cite{mozo}.  
\\

The author is grateful to Prof. S. Saito for the 
hospitality of Tokyo Metropolitan University. 

\end{document}